\documentclass{emulateapj}
\usepackage{natbib}
\def\mpchi{\,h^{-1}{\rm {Mpc}}}

\def\s{\mathbf{s}}
\def\be{\begin{equation}}
\def\ee{\end{equation}}
\def\ba{\begin{eqnarray}}
\def\ea{\end{eqnarray}}

\shorttitle{Fiber Collision Correction in Galaxy
Clustering} \shortauthors{Guo, Zehavi, \& Zheng}

\begin{document}

\title{A New Method to Correct for Fiber Collisions in Galaxy Two-Point Statistics}

\author{
Hong Guo\altaffilmark{1}, Idit Zehavi\altaffilmark{1}, Zheng
Zheng\altaffilmark{2}} \altaffiltext{1}{ Department of Astronomy, Case
Western Reserve University, 10900 Euclid Avenue, Cleveland, OH 44106}
\altaffiltext{2}{ Department of Physics and Astronomy, University of Utah,
115 South 1400 East, Salt Lake City, UT 84112 }

\begin{abstract}
In fiber-fed galaxy redshift surveys, the finite size of the fiber plugs
prevents two fibers from being placed too close to one another, limiting the
ability of studying galaxy clustering  on all scales. We present a new method
for correcting such fiber collision effects in galaxy clustering statistics
based on spectroscopic observations. The target galaxy sample is divided into
two distinct populations according to the targeting algorithm of fiber
placement, one free of fiber collisions and the other consisting of collided
galaxies. The clustering statistics are a combination of the contributions
from these two populations. Our method makes use of observations in tile
overlap regions to measure the contributions from the collided population,
and to therefore recover the full clustering statistics. The method is rooted
in solid theoretical ground and is tested extensively on mock galaxy
catalogs. We demonstrate that our method can well recover the projected and
the full three-dimensional redshift-space two-point correlation functions on
scales both below and above the fiber collision scale, superior to the
commonly used nearest neighbor and angular correction methods. We discuss
potential systematic effects in our method. The statistical correction
accuracy of our method is only limited by sample variance, which scales down
with (the square root of) the volume probed. For a sample similar to the
final SDSS-III BOSS galaxy sample, the statistical correction error is
expected to be at the level of 1\% on scales $\sim0.1$--$30\mpchi$ for the
two-point correlation functions. The systematic error only occurs on small
scales, caused by non-perfect correction of collision multiplets, and its
magnitude is expected to be smaller than 5\%. Our correction method, which
can be generalized to other clustering statistics as well, enables more
accurate measurements of full three-dimensional galaxy clustering on all
scales with galaxy redshift surveys.
\end{abstract}


\section{Introduction}
With fiber-fed spectrographs, galaxy surveys, such as the Las Campanas
Redshift Survey \citep{LCRS}, the 2dF Galaxy Redshift Survey (2dFGRS;
\citealt{2df}), the Sloan Digital Sky Survey (SDSS)
\citep{SDSSI,SDSSII,Strauss02}, and the Galaxy and Mass Assembly Survey(GAMA)
\citep{Driver11,Robotham10} can efficiently cover a large sky area and obtain
redshifts for a large set of targeted galaxies simultaneously. A well-known
problem of using fibers is that the finite size of the fiber plugs prevents
two fibers from being placed too close to one another on the same plate.
Consequently a significant fraction of targeted galaxies from a photometric
catalog cannot be assigned fibers and obtain measured spectroscopic
redshifts. This problem is partly alleviated by having some regions on the
sky covered by overlapping plates, but it still results in a fraction of
targeted galaxies left with no measured spectroscopic redshifts (e.g., $\sim
7\%$ in the SDSS). These fiber-collided galaxies are a hindrance to any
galaxy clustering study.  In this paper, we propose and test a new method to
account for fiber collision effects and to accurately measure galaxy
clustering statistics on small and intermediate scales.

The angular fiber-collision scale, under which two fibers on the same plate
collide with each other, is determined by the fiber placement hardware and
differs from survey to survey. For SDSS-I and II, this scale is $55\arcsec$,
corresponding to about $0.1\mpchi$ at the median redshift $z\sim0.1$. For
SDSS-III, the angular fiber-collision scale is slightly larger, $62\arcsec$,
corresponding to about $0.4\mpchi$ (comoving) at the larger median redshift
$z\sim0.55$ of the Baryon Oscillation Spectroscopic Survey (BOSS;
\citealt{SDSSIII,SDSS3,White11,DR9}). If fiber collisions are not corrected
for, galaxy clustering on scales below the collision scale would not be
accurately measured. For example, in the case of galaxy two-point correlation
function (2PCF), the effect is seen as a significant decline in the
clustering signal below the collision scale \citep{Jing98,Yoon08}. At a fixed
angular collision scale, the comoving scale increases with redshift, making
the fiber collision a more severe problem in studying small-scale clustering
for surveys at higher redshifts. Furthermore, fiber collisions have a
non-negligible effect on galaxy clustering measurements even on scales larger
than the collision scale \citep{Zehavi02,Zehavi05}. Therefore, to study
galaxy clustering on small scales and to have precise measurements on larger
scales, the effect induced by fiber collision has to be corrected for.

In general, there are two approaches to correct the fiber collision effect in
measuring galaxy 2PCFs. One is to recover the redshifts of the fiber-collided
galaxies and the other is to reconstruct the correct galaxy pair counts. For
the former one, a commonly adopted method is to assign each collided galaxy
the redshift of its nearest angular neighbor (nearest-neighbor correction;
e.g., \citealt{Zehavi02,Zehavi05,Berlind06}). The method has been applied to
measure the projected 2PCF $w_p(r_p)$ at a transverse separation $r_p$ for
the SDSS-I and II Main galaxy sample \citep{Zehavi02,Zehavi05,Zehavi10},
which proves to work well down to the fiber collision scale $\sim0.1\mpchi$.
However, the method fails below the collision scale and does not give a
satisfactory correction for the redshift-space correlation function. For
surveys at higher redshifts, the increase in the (comoving) fiber collision
scale limits the application of the nearest neighbor correction to clustering
at larger scales.

For the approach of reconstructing the pair counts, a common method is to
make use of the angular 2PCF of the spectroscopic sample, $w_{sz}(\theta)$,
and that of the parent photometric sample, $w_{pz}(\theta)$, at angular
separation $\theta$, with the same angular and redshift selection applied to
both samples \citep{Hawkins03,Li06a,Li06b,Ross08,White11}. The projected
galaxy pair count in computing $w_p(r_p)$ is weighted by the ratio
$F(\theta)=[1+w_{pz}(\theta)]/[1+w_{sz}(\theta)]$, where $\theta$ corresponds
to $r_p$ at the median redshift of the survey. Such a weighting scheme
incorporates the angular information of the missing galaxies and retrieves
the correct angular counts. However, $w_p(r_p)$ and $w_p(\theta)$ are not
completely correspondent. \cite{Li06b} test this method with mock catalogs of
SDSS galaxies. The method works reasonably well on large scales, but on
scales of $0.05$ --- $1\mpchi$ the corrected $w_p(r_p)$ shows a clear deficit
with respect to the true $w_p(r_p)$.

Other methods aimed at overcoming the fiber collision problem to measure the
small-scale galaxy clustering broadly fall into the above two categories or
some combination. For example, the small-scale clustering can be inferred
from the full photometric sample by cross-correlating with the spectroscopic
sample \citep{Eisenstein05,Masjedi06,Watson10,Watson11,Jiang11,Wang11}. Under
the assumption of isotropic clustering, the photometric objects near a
spectroscopic target can be considered to be at the same redshift as the
target, similar to the nearest neighbor method, and the small-scale
clustering can be measured. Contamination from interlopers reduces the
signal-to-noise and needs to be removed statistically. As with the nearest
neighbor method, this cross-correlation technique does not work in measuring
the full three-dimensional (3D) clustering in redshift space.

Accurate measurements of galaxy clustering are important in many
applications. In particular, the small-scale clustering can be used to probe
the spatial distribution of galaxies inside the host dark matter halos (e.g,
\citealt{Watson10,Watson11}), to infer the kinematics of galaxies in halos,
and to reveal environmental effects on galaxy formation and evolution. In
this paper we propose a new and efficient correction method for the fiber
collision effect based on the spectroscopic galaxy sample. The method is
proven to work well in the case of measuring both the projected galaxy 2PCFs
and the full 3D redshift-space 2PCFs. After describing the mock catalogs used
for testing the method and our measurements in Section 2, we present our
correction method and explain its theoretical basis in
Section~\ref{sec:methods}. Tests of the method with mock catalogs and
comparisons with other methods are presented in Section~\ref{sec:results}. We
discuss possible systematic effects of the method in Section~\ref{sec:sys}
and conclude in Section~\ref{sec:discussion}.

\section{Mock Catalogs and Clustering Measurements}
\label{sec:mocks}

Throughout the paper, our proposed new method will be tested and compared to
other methods using clustering measurements performed on available realistic
mock catalogs. We use the LasDamas galaxy mock
catalogs\footnote{http://lss.phy.vanderbilt.edu/lasdamas/} (C.\ McBride et
al. 2012, in prep.), which are constructed by populating galaxies into dark
matter halos identified in the LasDamas simulations. The $N$-body LasDamas
simulations adopt a spatially flat $\Lambda$CDM cosmology, with a matter
density parameter $\Omega_m=0.25$, baryon density of $\Omega_b=0.04$,
$\sigma_8=0.8$ (the primordial matter fluctuation amplitude on scales of
8$\mpchi$, linearly extrapolated to $z=0$), a primordial matter fluctuation
spectra index $n_s=1$, and a Hubble constant of $h=0.7$. Dark matter halos
are identified using a friends-of-friends algorithm (e.g., \citealt{Davis85})
with a linking length of $0.156$ times the mean particle separation. Dark
matter halos are populated with galaxies through a halo occupation
distribution (HOD) approach (e.g., \citealt{Berlind02,Cooray02,Zheng05}) and
the HOD parameters are determined through modeling the clustering of the
early BOSS $z\sim0.5$ CMASS sample \citep{White11}. Redshift space
distortions are also included in the mock catalogs by accounting for the
peculiar velocities of galaxies. The radial and angular selection functions
in the mock catalogs are constructed to be uniform.

In total, we make use of 40 LasDamas mock galaxy catalogs. In addition to
matching the clustering of CMASS galaxies, the mocks also reproduce the
geometry of the early BOSS data. As in \cite{White11}, the LasDamas mock
catalogs for CMASS sample are divided into three separate regions. For
simplicity, we only use the so-called ``region B" in \cite{White11}, since it
has the largest volume and number of galaxies. Each mock of this region
consists of about $50,000$ galaxies, with a sky coverage of $\sim 600\deg^2$
and a redshift range of $0.4 < z < 0.6$, corresponding to a volume of $\sim
0.16 h^{-3}{\rm Gpc}^3$. The mock catalogs do not have the specific tiling
mask and fiber collisions imposed on them, and we do that ourselves for our
tests as described below.

In this paper, we focus our discussion on the 2PCFs and related statistics.
We use the Landy-Szalay estimator \citep{Landy93} to measure the 2PCFs of
galaxies in the mock catalogs,
%%%%%%%%%%%%%%%%%%%%%%%%%%%%%%%%%%%%%%%%%%%%%%%%%%
\begin{equation}
\xi=\frac{DD-2DR+RR}{RR},\label{eqn:estimator}
\end{equation}
%%%%%%%%%%%%%%%%%%%%%%%%%%%%%%%%%%%%%%%%%%%%%%%%%%
where $DD$, $DR$ and $RR$ are the data-data, data-random, and random-random
pair counts measured from the data of $N$ galaxies and random samples
consisting of $N_R$ random points. These pair counts are normalized by
dividing by $N(N-1)/2$, $NN_R$, and $N_R(N_R-1)/2$, respectively.

We measure the 3D $\xi(r_p,\pi)$ and $\xi(s,\mu)$ functions and the
redshift-space 2PCF $\xi(s)$, where $r_p$ and $\pi$ are the separations of
galaxy pairs perpendicular and parallel to the line of sight
\citep{Fisher94,Zehavi02}, $s^2=r_p^2+\pi^2$, and $\mu=\pi/s$ is the cosine
of the angle between $\s$ and $\mathbf{\pi}$. The redshift-space 2PCF differs
from the real-space one because of the redshift distortion effect induced by
galaxy peculiar velocity. The redshift distortion can be mitigated by
projecting the 2PCF along the line-of-sight direction, with the projected
2PCF $w_p(r_p)$ \citep{Davis83} defined and measured as
%%%%%%%%%%%%%%%%%%%%%%%%%%%%%%%%%%%%%%%%%%%%%%%%%%
\begin{eqnarray}
w_p(r_p)=2\int_0^\infty \xi(r_p,\pi)d\pi
=2\sum_i\xi(r_{p},\pi_i)\Delta\pi_i, \label{eqn:wp}
\end{eqnarray}
%%%%%%%%%%%%%%%%%%%%%%%%%%%%%%%%%%%%%%%%%%%%%%%%%%
where $\pi_i$ and $\Delta\pi_i$ are the $i$th bin of the line-of-sight
separation and its corresponding bin size.

Following \cite{Hamilton92}, the redshift-space 2PCF $\xi(s,\mu)$ can be
written in the form of multipole expansion,
%decomposed as a sum of spherical harmonics,
%%%%%%%%%%%%%%%%%%%%%%%%%%%%%%%%%%%%%%%%%%%%%%%%%%
\begin{equation}
\xi(s,\mu)=\sum_l\xi_l(s)P_l(\mu),
\end{equation}
%%%%%%%%%%%%%%%%%%%%%%%%%%%%%%%%%%%%%%%%%%%%%%%%%%
where $P_l$ is the $l$-th order Legendre polynomial. The multipole moments
$\xi_l$ is determined by
%%%%%%%%%%%%%%%%%%%%%%%%%%%%%%%%%%%%%%%%%%%%%%%%%%
\begin{equation}
\xi_l(s)=\frac{2l+1}{2}\int_{-1}^{1}\xi(s,\mu)P_l(\mu)d\mu.
\label{eqn:pole}
\end{equation}
%%%%%%%%%%%%%%%%%%%%%%%%%%%%%%%%%%%%%%%%%%%%%%%%%%
In linear theory, only the moments of $l=0$, $2$, and $4$ are non-zero. The
monopole $\xi_0(s)$, quadrupole $\xi_2(s)$, and hexadecapole $\xi_4(s)$ are
useful for the study of redshift distortions and for obtaining constraints on
cosmological parameters \citep[see,
e.g.,][]{Hamilton92,Cole94,Tinker06,Padmanabhan08,Kazin11,Reid11}. We will
also test how well our method recovers these moments. %$\xi_0$ and $\xi_2$.

For measuring the 2PCFs, we use a binning scheme of $\Delta\log r_p=0.2$,
$\Delta\pi=1\mpchi$, $\Delta \log s=0.2$, and $\Delta\mu=0.1$. Our results
are insensitive to these choices. To obtain the projected 2PCF $w_p(r_p)$,
$\xi(r_p,\pi)$ is summed along the line-of-sight direction up to $\pi_{\rm
max}=40\mpchi$. Integrating to a larger line-of-sight separation or using
realistic angular and redshift selection functions will not affect the
conclusion on our correction method. We also note that while the mock
catalogs we use were constructed for studying the BOSS CMASS sample, our
general conclusions regarding the validity of our method are not dependent on
that and would hold for other mock and real data sets.

\section{The New Correction Method}
\label{sec:methods}

\subsection{Sample Division}
\label{sub:division}

In modern galaxy redshift surveys, a tiling algorithm is usually applied to
design and place spectroscopic plates (tiles) to cover the survey area. These
tiles partially overlap over some of the observed region. In such overlap
regions, a galaxy with no fiber assigned in one tile can have a fiber
allocated in the other one. For example, in SDSS-I and II, about $40\%$ of
the survey area is covered by more than one tile, which eliminates most of
the fiber collisions in those regions.

The basic idea of our method of dealing with the fiber collision effect is
simply to estimate the contribution of the fiber-collided galaxies to the
clustering by using the information in the tile overlap regions. In order to
make such an estimate, before measuring any clustering statistic, we divide
the full galaxy target sample (i.e., the input photometric sample) into two
distinct populations:

{\it Population 1}: a subsample in which each galaxy is not angularly
collided with any other galaxy in this subsample. We maximize the number of
galaxies that are not collided with each other. Such a set of ``decollided''
galaxies provide a ``clean'' subsample with no fiber collision correction to
be considered.

{\it Population 2}: a subsample including all the galaxies that are not in
Population 1. This is the set of potentially collided galaxies, and all the
fiber-collided galaxies come from this subsample. Each galaxy in this
population is within the fiber collision scale of a galaxy in Population 1.

The division of Population $1$ and $2$ follows the scheme of assigning fibers
in SDSS observations \citep{Blanton03}. The specific tiling algorithm always
recovers one galaxy from collided pairs, and two galaxies from the collided
triples if the angular distance of these two galaxies are larger than the
fiber collision scale. With our division, Population 1 galaxies always have
fibers allocated and have spectroscopic redshifts measured. Some fraction of
Population 2 galaxies can also have fibers allocated in the tile overlap
regions. Also with our division, in each ``collided'' close pair of galaxies,
one galaxy (with measured redshift) will always be part of Population $1$ and
the other (with a measured redshift or not) will be part of Population $2$.
This definition aims to ensure that the pair counts involving the galaxies
with redshifts in Population $2$ (hereafter, we refer to them as ``resolved''
galaxies) can be regarded as a representative subset of the overall pair
counts. The specific tiling and fiber assignment constraints can make the
situation non-trivial if the ``representative'' assumption is not satisfied
in real observation, and we discuss the possible systematics introduced in
such a case in Section \ref{sec:sys}.

We assume that in the survey, we have $N=N_1+N_2$ galaxy targets, where $N_1$
and $N_2$ are the numbers of galaxies in Populations $1$ and $2$,
respectively. Again, all galaxies in Population 1 have spectroscopic
redshifts, and because of fiber collisions only a fraction of Population 2
galaxies do (e.g., in tile overlap regions). We denote the set of targeted
galaxies as $D$ (with a total number $N$) and use $D_1$ and $D_2$ to
represent the sets of Population 1 and 2 galaxies in the sample (with numbers
of $N_1$ and $N_2$, respectively). We use $D^\prime$ and $D_2^\prime$ to
denote the corresponding galaxy sets that have spectroscopic redshifts
measured (with numbers $N^\prime$ and $N_2^\prime$). Note that by definition,
$D_1^\prime$ is the same as $D_1$.

The data-data pairs counts can then be decomposed as
%%%%%%%%%%%%%%%%%%%%%%%%%%%%%%%%%%%%%%%%%%%%%%%%%%
\begin{equation}
\label{eqn:pair_decomp_a} N_{DD}=N_{D_1D_1}+N_{D_1D_2}+N_{D_2D_2},
\end{equation}
and
\begin{equation}
\label{eqn:pair_decomp_b}
N_{D'D'}=N_{D_1D_1}+N_{D_1D'_2}+N_{D'_2D'_2}.
\end{equation}
%%%%%%%%%%%%%%%%%%%%%%%%%%%%%%%%%%%%%%%%%%%%%%%%%%
The actual numbers of pairs in equation~(\ref{eqn:pair_decomp_a}) are what
are needed to estimate galaxy 2PCFs, while the pair numbers in
equation~(\ref{eqn:pair_decomp_b}) are what one obtains in the spectroscopic
sample. Our method is to make use of the pair counts of Population 1 and 2
galaxies and those of Population 2 galaxies in the tile overlap regions to
recover the correct counts appearing in equation~(\ref{eqn:pair_decomp_a}),
and therefore to measure the 2PCFs properly. We note that the fraction of
$D_2'$ in $D_2$ galaxies, $N_2'/N_2$, is an important factor in the following
discussions. As with the data-data pair counts, the data-random pair counts
$DR$ can also be decomposed in a similar way.

\subsection{The Simplified Case}
\label{sub:ideal}

To illustrate our fiber-collision correction method, we first consider the
simplest case where Population 2 galaxies are randomly selected to be
assigned fibers. In this case, $D_2'$ is a random subset of $D_2$, and the
pair counts $N_{D_1D'_2}$ and $N_{D'_2D'_2}$ are simply proportional to the
full counts $N_{D_1D_2}$ and $N_{D_2D_2}$, i.e.,
%%%%%%%%%%%%%%%%%%%%%%%%%%%%%%%%%%%%%%%%%%%%%%%%%%
\begin{eqnarray}
\label{eqn:corrected_pair} N_{D_1D_2}={N_2\over
N_2'}N_{D_1D_2'},\quad N_{D_2D_2}={\left(N_2\over
N_2'\right)^2}N_{D_2'D_2'}.\label{eqn:d2d2}
\end{eqnarray}
%%%%%%%%%%%%%%%%%%%%%%%%%%%%%%%%%%%%%%%%%%%%%%%%%%
For simplicity, $N_2$ and $N'_2$ are assumed to be large so that $N_2(N_2-1)$
and $N'_2(N'_2-1)$ are replaced with $N_2^2$ and $N_2^{\prime 2}$,
respectively. The above equations provide the way to correct the pair counts
obtained from the spectroscopic sample.

Similarly, the data-random pair counts can be corrected as
%%%%%%%%%%%%%%%%%%%%%%%%%%%%%%%%%%%%%%%%%%%%%%%%%%
\begin{equation}
N_{D_2R}={N_2\over N_2'}N_{D_2'R}.
\end{equation}
%%%%%%%%%%%%%%%%%%%%%%%%%%%%%%%%%%%%%%%%%%%%%%%%%%
Since we can have a large number of random points, the correction here for
$N_{D_2R}$ is less noisy than those for $N_{D_1D_2}$ and $N_{D_2D_2}$.

The full pair counts are reconstructed as
%%%%%%%%%%%%%%%%%%%%%%%%%%%%%%%%%%%%%%%%%%%%%%%%%%
\begin{eqnarray}
N_{DD}&=&N_{D_1D_1}+{N_2\over N'_2}N_{D_1D'_2}+{\left(N_2\over
N'_2\right)^2}N_{D'_2D'_2}
\label{eqn:dd}\\
N_{DR}&=&N_{D_1R}+{N_2\over N'_2}N_{D'_2R}.\label{eqn:dr}
\end{eqnarray}
%%%%%%%%%%%%%%%%%%%%%%%%%%%%%%%%%%%%%%%%%%%%%%%%%%

\begin{figure}
\epsscale{1.2} \plotone{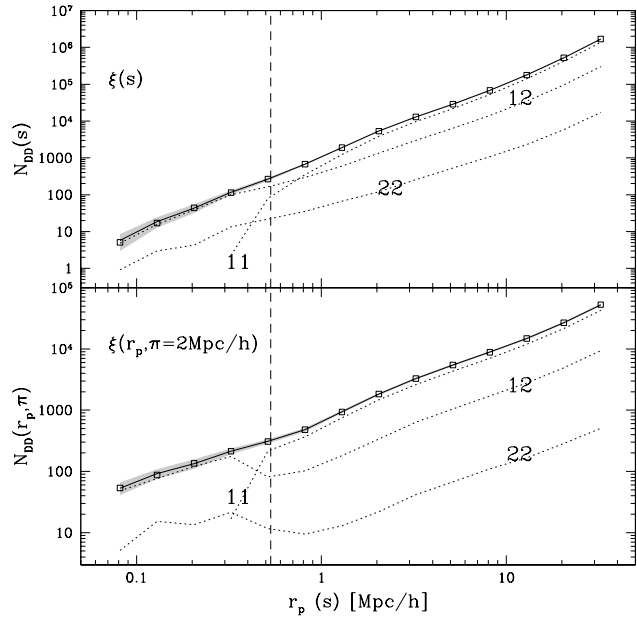} \caption{The full pair
counts $N_{DD}$ (solid lines) for $\xi(s)$ and $\xi(r_p,\pi)$ in the
simplified case. For the latter, we only show the example of $\pi=2\mpchi$.
The solid and dotted curves are the corrected pair counts averaged over
the 40 mocks, obtained from the spectroscopic galaxies, while the squares
show the actual pair counts. The
shaded areas are the $1\sigma$ error distribution of the corrected
$N_{DD}$. The corrected pair counts are further decomposed as the
contribution from the different populations, $N_{D_1D_1}$,
$N_{D_1D_2}$, and $N_{D_2D_2}$, shown as the dotted lines, labeled
here simply as $11$, $12$, and $22$. The vertical dashed lines denote the
physical fiber collision scale corresponding to the fiber
collision angular constraint, determined by the highest redshifts in the mocks.
} \label{fig:ratio}
\end{figure}

Figure~\ref{fig:ratio} shows the data-data pair counts from the mock catalogs
for the simplified case, decomposed according to equation~(\ref{eqn:dd}). The
fiber collisions are artificially imposed on the mocks with an overall
fraction of $N_2'/N_2=0.42$ as in the BOSS CMASS catalogs. The solid and
dotted curves are the corrected pair counts, obtained from the spectroscopic
samples, averaged over the 40 mocks, while the squares show the actual full
pair counts. The shaded region in each panel denotes the $1\sigma$ scatter
from the 40 mock catalogs of the $DD$ pair count (estimated from $D'D'$ based
on eq.~\ref{eqn:corrected_pair}). Comparing the solid curve (plus the small
shaded region) with the squares, we see that our correction method accurately
recovers the true pair counts over all scales probed, for both $\xi(s)$ and
$\xi(r_p,\pi)$. The increase in the scatter on small scales is caused by shot
noise, since the numbers of $D_1D_2'$ and $D'_2D'_2$ pairs are less than
$D_1D_2$ and $D_2D_2$ pairs.

From Figure~\ref{fig:ratio}, we see that there are almost no $D_1D_1$ pairs
with separation smaller than $\sim0.3\mpchi$, which is the minimum fiber
collision scale (corresponding to the lowest redshift $z=0.4$). On small
scales, the pair count is dominated by $D_1D_2$ pairs, while $D_2D_2$ pairs
have a small but non-negligible contribution. Across the fiber collision
scale ($\sim 0.3$--$0.5\mpchi$), the dominant contribution to the total pair
count shifts from $D_1D_2$ pairs to $D_1D_1$ pairs, reflecting the change
from collided to decollided galaxies.

\subsection{Theoretical Basis}
\label{theory}

Our correction method can be put in terms of a decomposition of the 2PCF.
With a galaxy sample divided into subsamples, such as red and blue galaxies,
central and satellite galaxies, or in our case Population 1 and 2 galaxies,
the 2PCF can be decomposed into contributions from the two-point auto- and
cross-correlation functions of subsample galaxies \citep{Zu08}. This is a
fully equivalent way of describing the pair decomposition. In the case of
dividing the galaxy sample into Population 1 and 2 subsamples, we have
%%%%%%%%%%%%%%%%%%%%%%%%%%%%%%%%%%%%%%%%%%%%%%%%%%
\begin{eqnarray}
N^2\xi&=&N_1^2\xi_{11}+2N_1N_2\xi_{12}+N_2^2\xi_{22}\label{eqn:xid}\\
&\approx&N_1^2\xi_{11}+2N_1N_2\xi_{12'}+N_2^2\xi_{2'2'},\label{eqn:xid'}
\end{eqnarray}
%%%%%%%%%%%%%%%%%%%%%%%%%%%%%%%%%%%%%%%%%%%%%%%%%%
where $\xi$ can be either $\xi(s)$, $\xi(r_p,\pi)$, or $w_p(r_p)$, and
$\xi_{11}$, $\xi_{12}$, and $\xi_{22}$ are the two-point auto-correlation
function of Population 1 galaxies, cross-correlation function between
Population 1 and Population 2 galaxies, and auto-correlation function of
Population 2 galaxies, respectively. The second line in the equation uses the
component correlation functions $\xi_{12'}$ and $\xi_{2'2'}$ estimated from
the $D_1$ and $D_2'$ galaxies in the spectroscopic sample to approximate
$\xi_{12}$ and $\xi_{22}$, which is the key point of our correction method.

If $D_2'$ galaxies are representative of $D_2$ galaxies, i.e., $D_2'$ is a
random subset of $D_2$, the ensemble average of $\xi_{12'}$ ($\xi_{2'2'}$)
should equal to that of $\xi_{12}$ ($\xi_{22}$),
%%%%%%%%%%%%%%%%%%%%%%%%%%%%%%%%%%%%%%%%%%%%%%%%%%
\be \langle\xi_{12'}\rangle=\langle\xi_{12}\rangle,\quad
\langle\xi_{2'2'}\rangle=\langle\xi_{22}\rangle. \ee
%%%%%%%%%%%%%%%%%%%%%%%%%%%%%%%%%%%%%%%%%%%%%%%%%%
That is, with ensemble averages, the approximate symbol can be replaced with
the equal sign. Therefore, our method using equation~(\ref{eqn:xid'})
provides an unbiased correction, which is a desired merit. In practice, the
method is only applied to one realization in the ensemble, so differences are
expected between the true underlying 2PCF and the 2PCF corrected with our
method. However, any discrepancies would only result from the fact that a
smaller number of galaxies are used to estimate the component 2PCFs. They
show up as sample variance, not systematic errors. Potential systematics
caused by the violation of the ``representative'' assumption of $D_2'$
galaxies are discussed in Section \ref{sec:sys}.

The component 2PCFs can be estimated with the Landy-Szalay estimator,
%%%%%%%%%%%%%%%%%%%%%%%%%%%%%%%%%%%%%%%%%%%%%%%%%%
\ba
\xi_{11}&=&{D_1D_1-2D_1R_1+R_1R_1\over R_1R_1},\label{eqn:xi11}\\
\xi_{12'}&=&{D_1D_2'-D_1R_2-D_2'R_1+R_1R_2\over R_1R_2},\label{eqn:xi12}\\
\xi_{2'2'}&=&{D_2'D_2'-2D_2'R_2+R_2R_2\over R_2R_2}\label{eqn:xi22}. \ea
%%%%%%%%%%%%%%%%%%%%%%%%%%%%%%%%%%%%%%%%%%%%%%%%%%
In the case that $D'_2$ is a random subset of $D_2$, we can use the same
random sample, i.e., $R_1=R_2=R$ in the above equations. It is not surprising
that when substituting the above equations into equation~(\ref{eqn:xid'}), we
end up with exactly the same result as with equations~(\ref{eqn:dd}) and
(\ref{eqn:dr}).

\subsection{The Tiled Case}
\label{sub:real}

Taking into account the tiling in the real observational situation would
change the geometry of the distribution of $D_2'$ galaxies such that most of
them would preferentially populate the tile-overlap regions. The survey
geometry can be described by the individual sectors, defined by areas of sky
covered by unique sets of tiles \citep{Blanton03}. We use $N_{\rm tile}$ to
denote the number of tiles covering each sector. For $N_{\rm tile}=1$
regions, essentially no fiber-collided galaxies ($D_2$) can have
spectroscopic redshifts. For $N_{\rm tile}\geq 2$ regions, however, most of
these will be resolved, with spectroscopic redshifts measured as a result of
repeated observations in these tile overlap regions. In practice, some
fiber-collided $D_2$ galaxies in $N_{\rm tile}=1$ regions may still have
spectroscopic redshifts from other surveys. In the case of the CMASS sample,
about $5.5\%$ of all the galaxies do not have redshifts. Specifically, the
values of the ratio $N_2'/N_2$ are about $5\%$, $71\%$ and $87\%$ in $N_{\rm
tile}=1, 2, 3$ regions, respectively. The $5\%$ recovered galaxies in $N_{\rm
tile}=1$ regions are included from the ``Legacy survey'' of SDSS-I and II
\citep{SDSS3}.

\begin{figure}
\epsscale{1.2} \plotone{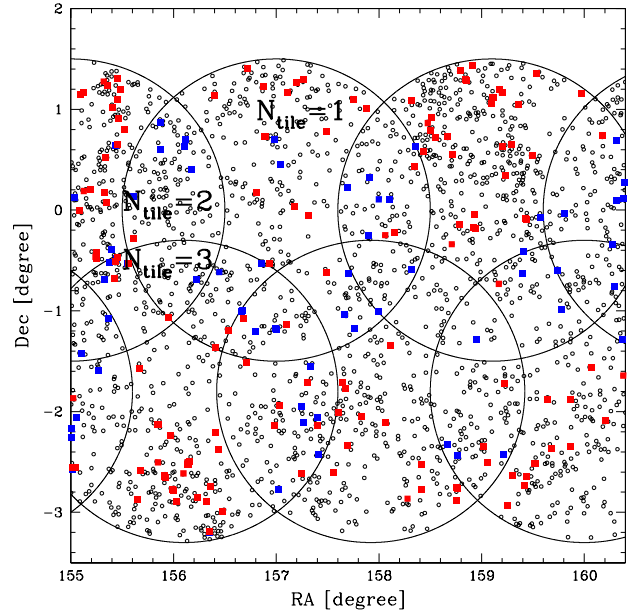} \caption{An example of the galaxy
distribution in one of our mocks. Large circles are the placed
plug-plate tiles. The open symbols are decollided $D_1$ galaxies. The blue
squares denote the $D_2'$ galaxies (i.e., resolved $D2$ galaxies, with fibers
assigned), while the red squares are
those $D_2$ galaxies without any fiber assigned. We also mark the
different $N_{\rm tile}$ regions in the figure.}
\label{fig:radec}
\end{figure}

In order to mimic the full observational case, we impose the tile placement
of the BOSS survey on all mock catalogs as well as fiber collisions according
to the tiling algorithm with the appropriate $N_2'/N_2$ ratios specified
above. We stress that the validity of our method does not depend on the
specific values of these parameters,  and these are simply chosen here to
resemble the values in the BOSS survey. Figure~\ref{fig:radec} illustrates
the galaxy distribution and the tile placement in a small section of one of
our mocks. The open symbols are all $D_1$ galaxies. Filled squares denote
$D_2$ galaxies, where blue ones mark resolved galaxies ($D_2'$) and red ones
are those galaxies in $D_2$ whose redshifts are missing due to fiber
collisions. The $D_2'$ galaxies are not randomly distributed over the whole
observed sky
--- they mostly occupy regions with $N_{\rm tile}\geq 2$.

For the tiled case, our correction method as in Section~\ref{theory} remains
the same, and the only modification is the need to account for the specific
geometry of the $D_2'$ galaxy distribution. One straightforward way to do
this is to create separate random catalogs $R_1$ and $R_2$ for $D_1$ and
$D_2'$ galaxies. The $R_1$ catalog can be created as usual by incorporating
the standard radial and angular selection functions of the sample. We note
that the latter is commonly characterized and applied as a function of the
individual sectors (e.g., \citealt{Zehavi02}). For the $R_2$ catalog, an
additional angular completeness mask needs to be applied, with the number of
random points in each individual sector modified by the corresponding
$N_2'/N_2$ ratio. For example, in Figure~\ref{fig:radec} the circular tile
centered at $RA\sim 158^\circ$ and $Dec\sim -2^\circ$ is comprised of 8
individual sectors, as a result of tile overlap, each with its own $N_2'/N_2$
value. In the extreme case that there are no $D_2'$ galaxies in $N_{\rm
tile}=1$ regions, these regions would be empty in the $R_2$ catalog.

Alternatively, one could choose to up-weight $D_2'$ galaxies in each sector
by $N_2/N_2'$, instead of creating a separate random sample for the $D_2'$
galaxies. However, for $N_{\rm tile}=1$ regions, $N_2'/N_2$ is usually a tiny
number, which will introduce large errors when adopting the up-weighting
method. Therefore, we prefer to down-weight the random catalog to accurately
account for the angular distribution and selection of $D_2'$ galaxies.

\section{Testing the Method}
\label{sec:results}

We test our correction method and compare with other commonly used methods by
measuring the 2PCFs and related statistics in the LasDamas mocks. We first
show results for the simplified case and then test our method on our tiled
mocks.

\subsection{The Simplified Case}
\label{sub:resim}

\begin{figure*}
\epsscale{1.0} \plotone{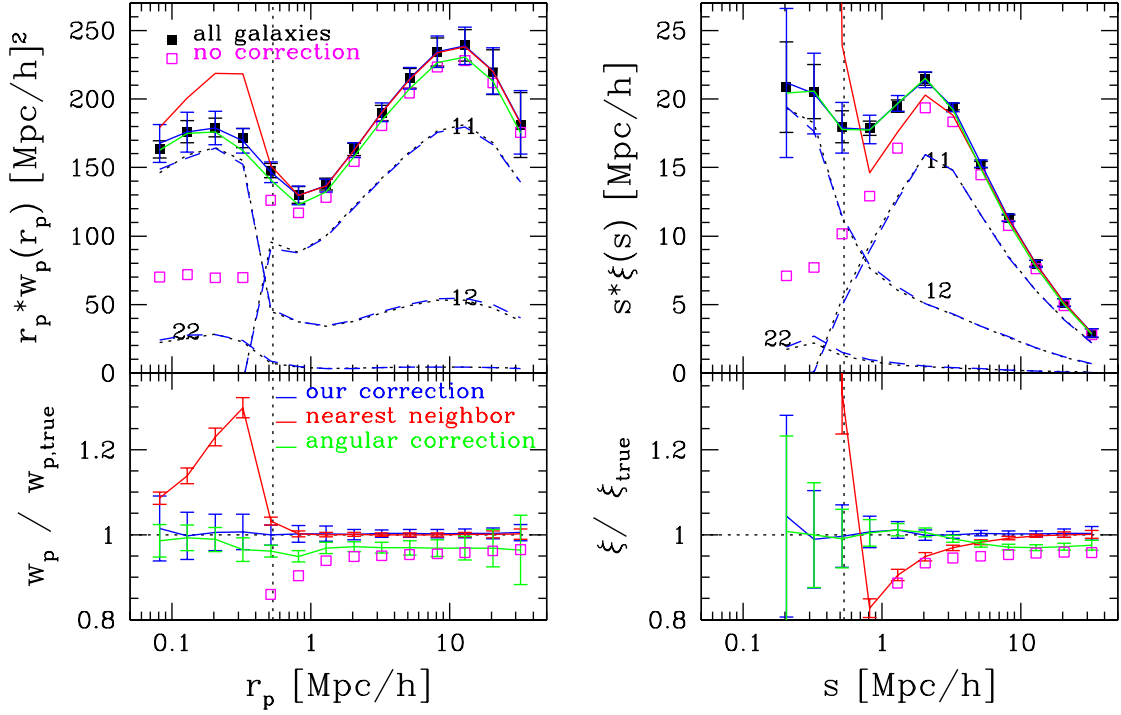} \caption{Tests of different
fiber-collision correction methods for the projected correlation function,
$w_p(r_p)$, and the redshift-space correlation function, $\xi(s)$, for the
simplified case. Top panels show the results for $r_p\times w_p(r_p)$ (left)
and $s\times\xi(s)$ (right) while the ratios of the estimates to the full
measurements, without any missing fiber-collided galaxies, are shown on the
bottom. Solid lines correspond to the different correction methods, filled
black squares are the true full measurements, and open magenta squares are
the results without any fiber-collision correction applied. Error bars
reflect the $1\sigma$ variation among the 40 LasDamas mock catalogs. (In top
panels, for clarify, error bars are only shown for the true case and for our
correction.)  We plot as well the individual components of the 2PCF
decomposition in the true case (black dotted lines; Eq.~\ref{eqn:xid}) and
for our correction (blue dashed lines; Eq.~\ref{eqn:xid'}). The vertical
dotted lines denote the physical fiber collision scale corresponding to the
fiber collision angular constraint, determined by the highest redshifts in
the mocks.} \label{fig:wp}
\end{figure*}

To apply our correction method, we first divide the mock galaxies into the
two populations as discussed in \S~\ref{sub:division}. We first define all
galaxies in the sample as $D_1$ and proceed as below: (1) For each collided
pair in $D_1$, we randomly change one galaxy to be $D_2$; (2) For each
collided triplet (and higher multiplet) in $D_1$, we assign as $D_2$ the
galaxy that collides with the most $D_1$ galaxies; (3) We repeat steps (1)
and (2) until none of the galaxies in $D_1$ collides with one another. This
method maximizes the number of galaxies in $D_1$ and mimics the real tiling
algorithm. In reality, some objects (e.g., quasars) have higher fiber
allocation priority and some assigned fibers do not result in reliable
redshifts due to the hardware limitation. This complicates the tiling
algorithm, but the targeted galaxies can still be divided into the two
populations properly. The population division results in about 9.5\% of all
galaxies being in the $D_2$ sample. We then randomly assign fibers to 42\% of
the $D_2$ galaxies (i.e., $N_2'/N_2=0.42$), so that about 5.5\% of {\it all}
galaxies do not have fibers assigned, mimicking the fiber collision fraction
in the CMASS sample. With the simplest case considered here, we only need to
create one random catalog.

Figure~\ref{fig:wp} shows the 2PCFs ($w_p$ and $\xi(s)$) from our correction
method and the comparisons with those from other methods. Filled squares
represent the actual 2PCFs, measured using all the mock galaxies. Error bars
reflect the $1\sigma$ variation among the 40 LasDamas mocks. Open squares are
obtained using only those galaxies that have fibers assigned, corresponding
to the case without any fiber-collision correction. In each panel, the
vertical dotted line is the fiber-collision scale corresponding to the
highest redshifts included in the mocks. Not accounting for fiber-collisions,
the measured 2PCF drops sharply below the collision scale, and it is also
underestimated on larger scales. Note that the error bars in the top panels
represent the variance (fluctuation) among the 40 mocks, and they do not
reflect directly the accuracy of our correction method. In the bottom left
panel, we compute the ratio $w_p/w_{p,{\rm true}}$ (and similarly for
$\xi(s)$ in the bottom right panel) for each mock, where $w_p$ and $w_{p,{\rm
true}}$ are the corrected 2PCF and the true one. The error bars are the
$1\sigma$ variation of the ratio among the 40 mocks, which reflects the mean
accuracy of our correction method for one mock. The mean ratio curve
corresponds to a volume 40 times larger than one mock, showing clearly that
our correction method is unbiased.

In each panel, the blue solid curve is the 2PCF obtained with our correction
method, averaged over the 40 mocks. Error bars are the $1\sigma$ scatter
among the 40 mocks. Our correction method works very well for both $w_p$ and
$\xi(s)$ over all measured scales. In particular, it works down to the
smallest scales for which we have enough pair counts to estimate the 2PCF
($r_p\sim0.1\mpchi$). The fractional errors for $w_p(r_p)$ on small scales
are smaller than those for $\xi(s)$, since for the same value and bin size of
$r_p$ and $s$ there are more pairs in computing $w_p(r_p)$ than $\xi(s)$ due
to the projection over large line-of-sight separation. On scales above
$1\mpchi$, the errors in both $w_p$ and $\xi(s)$ are similar, at the level of
$<$3\%.

The full collision-free 2PCFs and the ones corresponding to our new
correction method are further decomposed, in the top panels, into the
contributions from the component 2PCFs (dashed and dotted lines; see
Eq.~(\ref{eqn:xid}) and (\ref{eqn:xid'})). Our correction recovers well all
the components of the correlation function. Similar to
Figure~\ref{fig:ratio}, the contribution from $\xi_{11}$ dominates on large
scales, and it quickly decreases below the fiber collision scale. The
contribution from component $\xi_{12}$ dominates on scales smaller than the
collision scale, where that of $\xi_{22}$ makes a non-negligible
contribution. Without any fiber collision correction, the measured 2PCFs
(open squares) are still a combination of the three component 2PCFs, but with
different coefficients than those in equation~(\ref{eqn:xid'}). The
significant decline at the fiber collision scale in the non-corrected
$w_p(r_p)$ reflects the transition from $\xi_{11}$ to $\xi_{12}$ dominated
regime. The low amplitude below the collision scale suggests the lack of
contribution from $\xi_{12}$.

For comparison, the nearest-neighbor method (red curves) yields the correct
$w_p(r_p)$ above the fiber collision scale, due to the line-of-sight
projection, but it clearly overestimates $w_p(r_p)$ below this scale (see
also \citealt{Zehavi02,Zehavi05}), because of the increasing importance of
the line-of-sight separation on small scales. For $\xi(s)$, the
nearest-neighbor estimate fails to recover the correct values over most
scales, deviating by more than $1\sigma$ for all scales $<10\mpchi$. Thus its
use is very limited for redshift-space clustering measurements. The 2PCFs
(both $w_p$ and $\xi(s)$) from the angular correction method systematically
deviate from the true values except for small scales, and on large scales
they approach the estimate with no fiber-collision correction applied (see
\citealt{White11}). The deviation is at a level $<10\%$, consistent with the
finding of \cite{Li06b}. Physical explanations for the deviations seen in the
nearest-neighbor and angular corrections are provided in Section
\ref{sec:discussion}.

In contrast, our correction method appears unbiased, not showing any
systematic errors. The measurement errors are larger than those with all
galaxies (top panels in Figure~\ref{fig:wp}). This is easy to understand ---
our method only uses galaxies with fibers assigned (i.e., with spectroscopic
redshifts), which is less than the total number of galaxies and therefore the
sample variance increases. A survey of larger volume would help to reduce the
sample variance. The other two commonly used correction methods do introduce
scale-dependent systematic errors, although the nearest-neighbor method does
work remarkably well for $w_p$ measurements on large scales.

\subsection{The Tiled Case}
\label{sec:sub_tile}

\begin{figure*}
\epsscale{1.0} \plotone{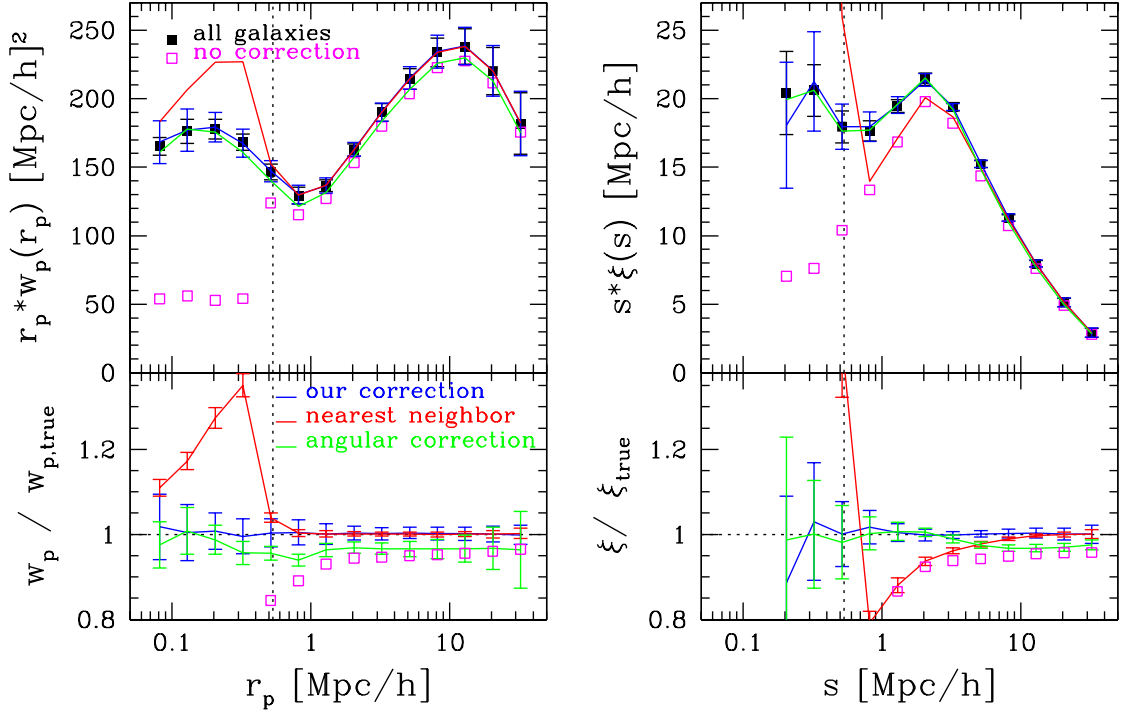} \caption{The same as Figure
\ref{fig:wp} but now for the tiled mocks. } \label{fig:wptile}
\end{figure*}

To test our method for the more realistic situation, we impose the BOSS
tiling geometry on the mocks (see an illustration in Figure~\ref{fig:radec}).
The division of $D_1$ and $D_2$ galaxies is still the same as in
section~\ref{sub:resim}. By definition, all $D_1$ galaxies have fibers
assigned. We then randomly assign fibers to $N_2'/N_2=$ 5\%, 71\%, and 87\%
of $D_2$ galaxies in $N_{\rm tile}=$ 1, 2, and 3 regions, respectively. These
fractions are consistent with those in the CMASS sample. As with the simplest
case, there are about 5.5\% of {\it all} galaxies left without fibers.

As explained in section~\ref{sub:real}, we create separate random catalogs
$R_1$ and $R_2$ for $D_1$ and $D_2'$ galaxies, respectively. The $R_1$
catalog has the overall survey geometry of the mock samples. For the $R_2$
catalog, in addition to the overall geometry, we impose a random subsampling
of a  $N_2'/N_2$ fraction of points in each individual sector.  Note that
when applying the method on actual data, it's best to use the individual
$N_2'/N_2$ values in each sector for constructing $R_2$, rather than the
average value for each $N_{\rm tile}$, to fully account for the distribution
of $D_2'$ galaxies.

The 2PCFs results from the different correction methods on the tiled mocks
are shown in Figure~\ref{fig:wptile}. They are quite similar to those seen in
Figure~\ref{fig:wp} for the simplified case. Our correction method still
accurately recovers both $w_p(r_p)$ and $\xi(s)$ over all measured scales.
The errors increase only slightly, compared to the previous case, reflecting
the increased sample variance (as $D_2'$ galaxies now occupy mostly the
smaller tile overlap regions). For $w_p(r_p)$, the errors are about $8\%$ at
$r_p\sim0.1\mpchi$ and $13\%$ at $r_p\sim30\mpchi$ for one mock (top left
panel in Figure~\ref{fig:wptile}). These errors include the intrinsic
fluctuation among the mocks. Similar to Figure~\ref{fig:wp}, the bottom
panels of Figure~\ref{fig:wptile} plot the ratios of $w_p/w_{p,{\rm true}}$
and $\xi/\xi_{\rm true}$, reflecting the expected accuracy of the correction
methods. For $w_p$,  the correction errors for our method are about $6\%$ at
$r_p\sim0.1\mpchi$ and $\sim2.5\%$ at $r_p\sim30\mpchi$ for the volume of one
mock. The mean ratio in each panel again demonstrates that the method is
unbiased. The success of our method also implies that applying the
completeness $N_2'/N_2$ on the $R_2$ random catalog is the proper way to
account for the angular distribution of $D_2'$ galaxies.  The level of
accuracy and associated systematics  for the nearest-neighbor and angular
corrections also remain similar to those in the simplified case.

\begin{figure*}
\epsscale{1.0} \plotone{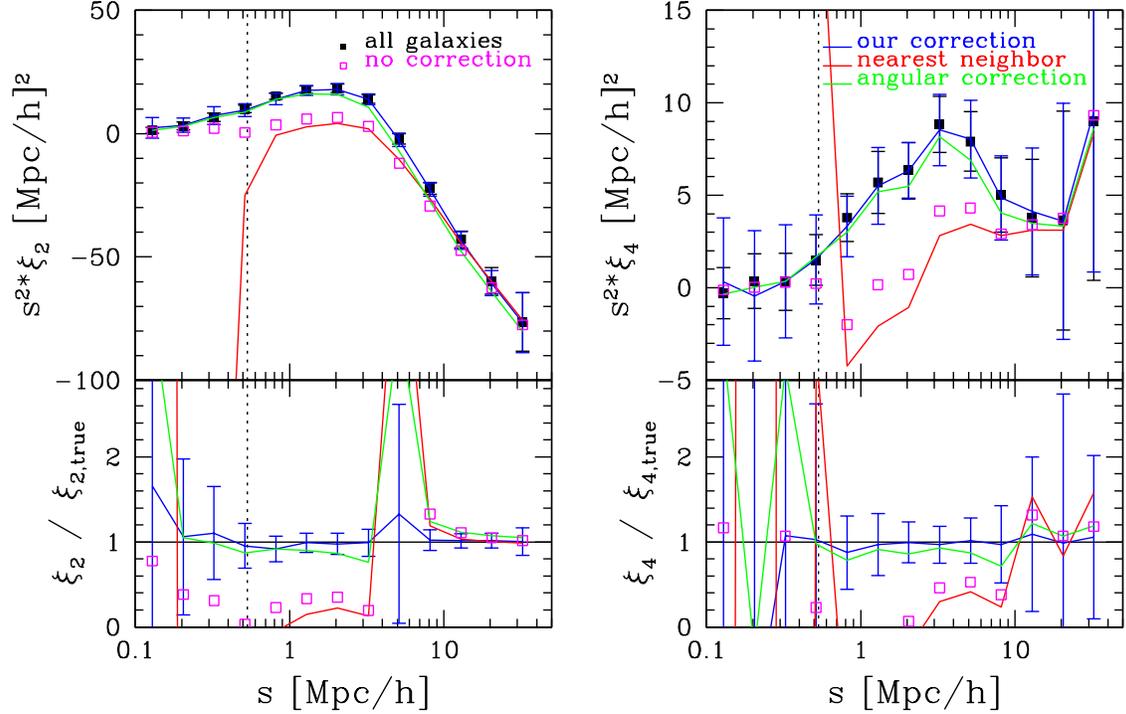} \caption{The quadrupole $\xi_2(s)$ (left
panel) and the hexadecapole $\xi_4(s)$ (right panel) from different
correction methods. The lines and symbols are similar to Figure
\ref{fig:wptile}. } \label{fig:poles}
\end{figure*}

We have tested as well on the mocks the recovery of the full 3D
$\xi(r_p,\pi)$ correlation function and its moments. The 2PCF $\xi(s)$ in the
right panels of Figure~\ref{fig:wptile} is the monopole of the 3D redshift
space 2PCF. In Figure~\ref{fig:poles}, we show the quadrupole $\xi_2(s)$
(left) and hexadecapole $\xi_4(s)$ (right) from the different correction
methods. Again, our method provides unbiased estimates of these quantities.
Note that the large error bars in the ratios near $s\sim 0.1\mpchi$ and
$s\sim5\mpchi$ in the bottom left panel and below $s\sim 0.4\mpchi$ in the
bottom right panel result from the fact that the corresponding multipole is
near zero at these scales. The success of our method has important
implications on studying redshift-space distortions in the non-linear regime
(e.g., \citealt{Tinker06,Tinker07,Reid11}), which has been hindered so far by
the fiber collision effect. In contrast, the nearest neighbor method
completely fails to recover the small-scale redshift distortions, because it
effectively reduces the line-of-sight separations of pairs, washing out the
Fingers-of-God signal. Residual systematics remain for it on larger scales as
well. The angular correction method works reasonably well, but systematic
deviations at a 10\% level persist in the quadrupole and hexadecapole.

\section{Systematic Effects}
\label{sec:sys}

On key assumption in our method presented above is that the $D_2'$ galaxies
are a representative (or random) subset of $D_2$ galaxies, which ensures that
pair counts involving $D_2$ galaxies can be recovered from those involving
$D_2'$ galaxies in a simple way.

In reality, however, the ``representative'' assumption may not be fully
satisfied, given the constraints from tiling and fiber assignment algorithm.
There are two types of potential systematic effects in our correction method,
when applied to the real observational data. The first is related to the
possible difference in the target density between the overlap and non-overlap
regions, due to specifics of the tiling algorithm. The second is related to
galaxy pairs in collided triplets (or higher-order collided groups). In what
follows, we discuss these two effects and provide solutions.

\subsection{Density Effect}

If the tiling algorithm is optimized to assign the most fibers to the galaxy
targets, it may preferentially place overlapping tiles in higher number
density regions (\citealt{Blanton03}; Tinker et al. 2011, private
communication). In such a case, $D_2'$ galaxies come from regions of slightly
higher number density, not necessarily representative of the overall $D_2$
galaxies. This may limit the accuracy of our use of
$\langle\xi_{12'}\rangle=\langle\xi_{12}\rangle$ and
$\langle\xi_{2'2'}\rangle=\langle\xi_{22}\rangle$. Observationally, the
impact of such potential density variations on our method can be evaluated,
if we have a complete representative observed area, where all fiber-collided
galaxies have been resolved by repeat observations. Theoretically, a set of
mocks with the actual tiling algorithm used in observations applied directly
to it, can help understand the impact. Since neither the required observed
area nor the realistic tiled mocks are yet available, we perform various
tests with our mocks to estimate the impact of the density effect.

We define a galaxy density measure of the overlap regions (``overlap
density'') as $\delta\equiv n_{\rm overlap}/n_{\rm all}-1$, where $n_{\rm
overlap}$ and $n_{\rm all}$ are the number densities of galaxies in the
overlap regions and the whole survey regions, respectively. Then the question
becomes whether the accuracy of our recovered 2PCFs depends on the value of
$\delta$.  While we do not expect systematic density variations in the
overlap regions in the ensemble of our tiled mocks, the variation caused by
sample variance in individual mocks can be used to study the density effect.

\begin{figure}
\epsscale{1.2} \plotone{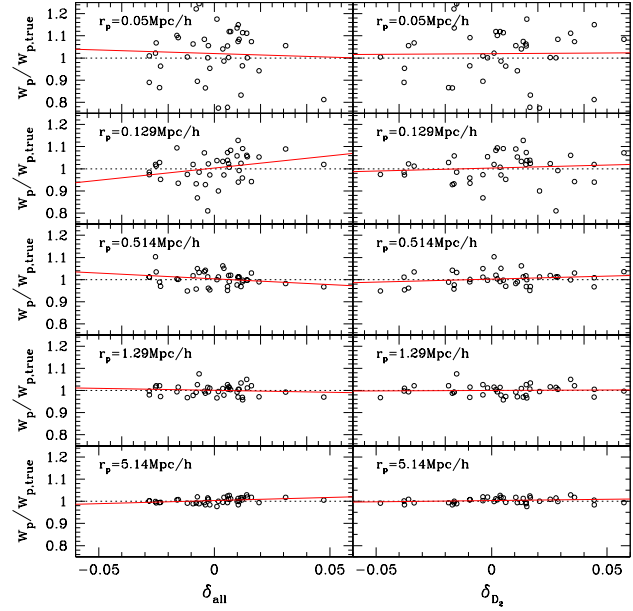}
\caption{
Dependence of the ratio $w_p/w_{p,{\rm true}}$ on the over-density of all
galaxies $\delta_{\rm all}$ and that of $D_2$ galaxies $\delta_{D_2}$
in tile overlap regions at various scales $r_p$. In each panel, the open
circles are the measurements from the $40$ LasDamas mocks and the solid
line is the linear least square fit to the data.}
\label{fig:delta}
\end{figure}

We have calculated the ``overlap density'' for each of our 40 mock catalogs
using the number densities of both all the galaxies and just the $D_2$ ones,
which we term as $\delta_{\rm all}$ and $\delta_{D_2}$, respectively.
Figure~\ref{fig:delta} shows the dependence of $w_p/w_{p,{\rm true}}$ on the
two overlap density measures, at different scales $r_p$. In each panel, the
open circles are the measurements from the $40$ LasDamas mocks and the solid
line is the linear least square fit to the data points. On large scales
($r_p\gtrsim 1\mpchi$), we find no dependence of the 2PCF on the overlap
density. On small scales ($r_p\lesssim 1\mpchi$), the scatter is large, but
the 2PCF ratio is again consistent with no dependence on the overlap density.
Overall, no strong systematic dependence on either $\delta_{\rm all}$ or
$\delta_{D_2}$ is found over the overlap density range probed by our mocks
($-0.06$ to $0.06$). To put this into the context of real observation, we
calculate the overlap densities in the CMASS DR9 sample and find them to be
$\delta_{\rm all}=0.032$ and $\delta_{D_2}=0.085$. Based on our tests, we
expect that the density effect in such a sample is not a concern. We have
also verified that the angular 2PCFs of the overlap and non-overlap regions
show consistent clustering amplitudes in the CMASS DR9 sample, confirming
that the density effect can be likely ignored.

The density effect is not significant in our mocks, but there may exist
another density-caused subtle effect in observational data. If $D_2'$
galaxies in the overlap regions and the fiber-collided $D_2$ galaxies are not
exactly the same type of galaxies given the environment difference, they
might have a systematic difference in clustering. We use the CMASS DR9 sample
to test this and find that the $D_2'$ and $D_2$ galaxies do have the same
color and apparent magnitude distribution, not unexpected given that the
tiling algorithm itself is just an angular selection and does not involve any
physical properties of the targeting galaxies.

As a whole, our investigation suggests that the density difference between
the tile overlap regions and the whole survey region does not introduce a
noticeable systematic effect in our method.

\subsection{Effect of Collision Groups}

There is another effect that can violate the ``representative'' assumption.
Galaxies in the $D_2$ population can be associated with different collision
groups. For $D_2$ galaxies that are part of collision pairs only (N.B. all
such pairs are $D_1$-$D_2$ or 1-2 pairs given that the number of $D_1$
galaxies is maximized in our division), $D_2'$ galaxies in overlap regions
are certainly a random subset of such $D_2$ galaxies. There are also $D_2$
galaxies that are part of collision triplets where the other two galaxies of
each triplet are part of $D_1$. For such $D_2$ galaxies in $D_1$-$D_2$-$D_1$
(1-2-1) triplets, $D_2'$ galaxies in overlap regions are also a random subset
of such $D_2$ galaxies. So the count of pairs including such $D_2$ galaxies
can be well reproduced with $D_2'$ galaxies in overlap regions, for the above
two cases. However, for $D_2$ galaxies in collision groups, where more than
two such galaxies of each group collide with each other, these $D_2D_2$
galaxy pairs would not be appropriately recovered in most overlap regions.
Since galaxy triplets are expected to dominate over higher multiplets, we
focus our discussion here on them.

Again, given that we maximize the number of $D_1$ galaxies, the collision
triplets are either 1-2-1 type or 1-2-2 type. The latter kind can not be
fully recovered in $N_{\rm tile}=2$ regions (only in $N_{\rm tile}>2$
regions) and is thus not appropriately represented when measuring the
clustering. Moreover, these $D_2$ galaxies might be clustered differently
than those in the previous two cases (1-2 pairs and 1-2-1 triplets). In our
mocks, the fraction of $D_2$ galaxies in 1-2 pairs, 1-2-1 triplets, and 1-2-2
triplets are $70\%$, $6\%$ and $22\%$, respectively. The collided galaxies in
such 1-2-2 triplets thus make up only about $1.2\%$ (22\% of 5.5\%) of the
total number of galaxies. This fraction is even smaller, about $0.7\%$, in
the CMASS DR9 sample. However, they can still have an adverse effect on the
clustering measurements.

In order to assess this, we  modify the way of assigning fibers to the
galaxies in our mocks to mimic the real tiling algorithm. In Section
\ref{sec:sub_tile}, the recovered $D_2$ galaxies in each sector are randomly
selected according to the prescribed fraction. We now assign fibers to $D_2$
galaxies in each colliding group according to the number of tiles covering
that sector. For example, for three galaxies colliding together in $N_{\rm
tile}=2$ regions, at most two of them would be assigned fibers, one from
$D_1$ and the other from $D_2'$.

\begin{figure*}
\epsscale{1.0} \plotone{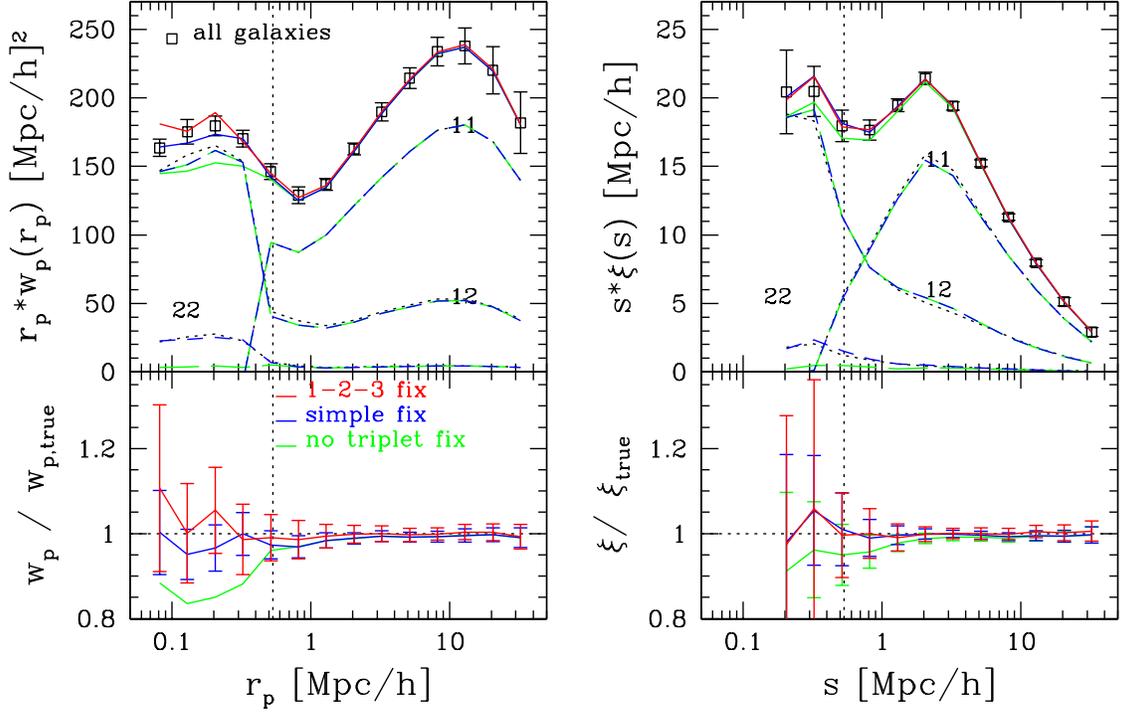} \caption{Effect of collision triplets
on $w_p(r_p)$ and $\xi(s)$, and comparison between the two correction
methods. The unresolved $D_2$ galaxies in collision triplets, if not
corrected, lead to an underestimate of the 2PCF on small scales (green). The
``1-2-3'' fix (red) and the simple fix (blue) methods can help to reduce the
effect, and the latter has smaller sample variance. In the top panels,
individual components of the 2PCF decomposition are plotted for the true case
(black dotted lines) and for our corrections (blue and green dashed lines). }
\label{fig:cpfix}
\end{figure*}

Figure \ref{fig:cpfix} shows the 2PCFs and the decomposition components for
our method without any correction for such a ``triplet effect'' (green lines)
compared with the true ones(open symbols and black dotted lines). It clearly
shows that, in this case, due to the missing $D_2D_2$ pairs the $\xi_{22}$
components on scales less than the fiber collision scale are significantly
decreased. The $\xi_{22}$ term has around a $15\%$ contribution to $w_p$ on
small scales, so without any correction for this the resulting $w_p$ is now
significantly underestimated below the fiber collision scale, and $\xi(s)$ is
also affected.

A natural solution to correct for this systematic effect is to extend
Equation (\ref{eqn:xid}) to three populations---$D_1$, $D_2$ and $D_3$, where
$D_2$ here refers to the population of galaxies that collide with $D_1$ but
not collide with each other, and $D_3$ is the rest of the galaxies,
corresponding to higher-rank collision groups. Equations (\ref{eqn:xid}) and
(\ref{eqn:xid'}) are then revised as,
%%%%%%%%%%%%%%%%%%%%%%%%%%%%%%%%%%%%%%%%%%%%%%%%%%
\begin{eqnarray}
N^2\xi&=&N_1^2\xi_{11}+N_2^2\xi_{22}+N_3^2\xi_{33}+2N_1N_2\xi_{12}\nonumber\\
&&+2N_1N_3\xi_{13}+2N_2N_3\xi_{23}\label{eqn:xid3}\\
&\approx&N_1^2\xi_{11}+N_2^2\xi_{2'2'}+N_3^2\xi_{3'3'}+2N_1N_2\xi_{12'}\nonumber\\
&&+2N_1N_3\xi_{13'}+2N_2N_3\xi_{2'3'}.\label{eqn:xid3'}
\end{eqnarray}
%%%%%%%%%%%%%%%%%%%%%%%%%%%%%%%%%%%%%%%%%%%%%%%%%%
We refer to this correction method as ``1-2-3 fix'' below. The limitation of
this correction is the small number of resolved $D_3'$ galaxies, since one
needs regions covered by three or more tiles to recover $D_3$ galaxies, which
only occupy a small fraction of the full survey region. Therefore such a
correction method would have large sample variance.

Another way of correcting for the collision triplets is to still use Equation
(\ref{eqn:xid'}) but simply add to $D_2'D_2'$ the estimated missing pair
counts in the tile overlap regions. The missing 2-2 pairs in 1-2-2 triplets
are statistically equivalent to 1-2 pairs in such triplets. We therefore
account for the unrepresented $D_2'D_2'$ close pairs using the recovered
$D_1D_2'$ pairs in these triplets. We denote the $D_2$ galaxies (the
definition of $D_2$ population here still follows Equation~\ref{eqn:xid}) in
1-2-2 collision triplets as $D_t$ and the recovered $D_2$ galaxies in such
triplets as $D_t'$. The total number of $D_2D_2$ close pairs in each
$(r_p,\pi)$ bin below the fiber collision scale {\it in overlap regions} can
then be written as follows:
%%%%%%%%%%%%%%%%%%%%%%%%%%%%%%%%%%%%%%%%%%%%%%%%%%
\begin{equation}
N_{D_2D_2}=
\left({N_{2}^{ol}\over {N_{2}'^{ol}}}\right)^2N_{D_2'D_2'}+{1\over2}\frac{N_t}{N_t'}N_{D_1D_t'}
\label{eqn:tripfix}
\end{equation}
\begin{equation}
=\left({N_{2}^{ol}\over {N_{2}'^{ol}}}\right)^2 \left[
N_{D_2'D_2'}+{1\over2}\frac{N_t}{N_t'}N_{D_1D_t'}\left({{N_{2}'^{ol}}\over {N_{2}^{ol}}}\right)^2\right] ,
\label{eqn:tripletfix}
\end{equation}
%%%%%%%%%%%%%%%%%%%%%%%%%%%%%%%%%%%%%%%%%%%%%%%%%%
where $N_t$ and $N_t'$ are the numbers of $D_t$ and $D_t'$ galaxies in the
tile overlap regions, and $N_{2}^{ol}$ and $N_{2}'^{ol}$ are the total and
recovered number of $D_2$ galaxies in overlap regions, respectively. The
first term in Equation~(\ref{eqn:tripfix}) represents the resolved $D_2D_2$
pairs in the sample (typically only in $N_{\rm tile}>2$ regions), while the
second term is an estimate of the missing close pairs (mostly in $N_{\rm
tile}=2$ regions). The expression $(N_t/N_t')N_{D_1D_t'}$ is the total number
of 1-2 pairs in 1-2-2 collision triplets, and dividing it by two gives the
expected number of 2-2 pairs in such triplets (since a 1-2-2 triplet has two
1-2 sides and one 2-2 side).  Note that for most triplets in overlap regions,
$N_t/N_t'=2$, so this simply adds a missing $D_2'D_2'$ pair for each
$D_1D_t'$ pair. The correction term itself is then the second term in the
square brackets in Equation~(\ref{eqn:tripletfix}), adding the normalized
missing pairs in 1-2-2 triplets below the fiber collision scale, and we refer
to it as the ``simple fix''.

Figure \ref{fig:cpfix} also shows the results of applying these two methods
to the mocks with the modified fiber assignment. Both corrections alleviate
most of the systematics. The ``1-2-3'' fix is in better agreement with the
underlying measurements (in particular for $\xi(s)$), but as discussed above,
it has large sample variance, leading to a $\sim$50\% increase in the error
bars. The simple fix reduces the bulk of the systematics, but a few percent
deficit remain on small scales. This could be caused by the complicated
structure of the high-order colliding groups, remaining systematics
associated with the $D_1D_2'$ pair counts and sample variance.

Although the two correction methods proposed here are not as accurate as in
the simple tiled case, they still provide workable estimates of the true
2PCFs. We regard the simple fix as the more practical one, due to its simple
application and the increased sample variance of the ``1-2-3'' fix. We thus
advocate simply incorporating the additional term
(Equation~\ref{eqn:tripletfix}) into the overall method. We have explored
many alternative corrections to the issue of collision groups, with varying
complexity and success, and will study it farther when a more realistic set
of tiled mocks or a complete fiber-collision free subsample in the BOSS
ancillary program becomes available. We note that, as mentioned above, the
LasDamas mocks have in fact $\sim60\%$ more close triplets than the real
CMASS sample. Thus the magnitude of the effect presented here and any
residual systematics are conservative and are likely smaller in the real
data.

\begin{figure*}
\epsscale{1.0} \plotone{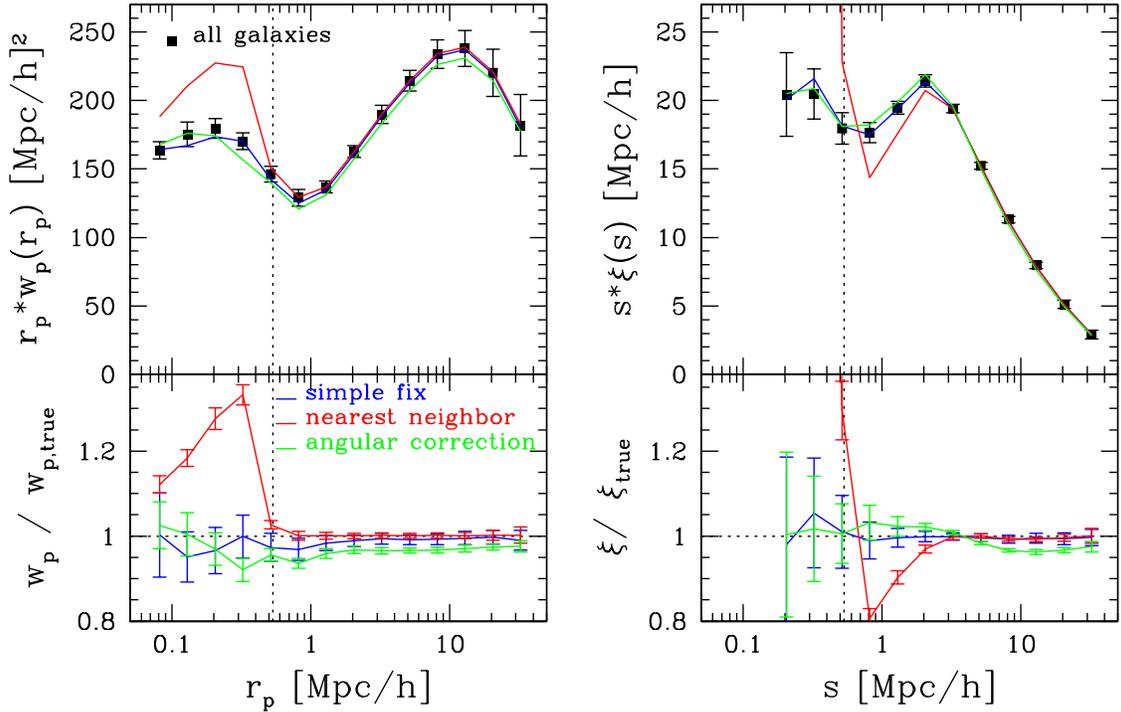} \caption{Effect of collision triplets
on the nearest neighbor and angular correction methods. The lines are the
same as in Figure \ref{fig:wptile}, but we adopt the simple fix as our
correction.} \label{fig:fix_com}
\end{figure*}
We also test the influence of the triplet effect in the mocks on the nearest
neighbor and angular correction methods, as shown in Figure
\ref{fig:fix_com}. The nearest neighbor method is mostly unaffected by this,
still recovering the correct correlation function on large scales. The
angular correction is impacted by the triplet effect, especially for the
redshift-space correlation function, though to a lesser extent than our
method. Based on our tests, the overall accuracy of our method appears to
still be better than the angular correction and nearest neighbor methods.

\section{Summary and Discussion}
\label{sec:discussion}

In this paper, we present a novel method for correcting the effects of fiber
collisions in galaxy clustering statistics, utilizing resolved fiber
collisions in tile overlap regions. The key element is dividing the target
galaxy sample into two distinct populations influenced differently by the
fiber collisions and combine their contributions according to equation
(\ref{eqn:xid'}). The collided galaxies, making up our so-called Population 2
subsample, are partially resolved, mainly in the tile overlap regions. These
resolved collided galaxies allow us to recover and measure the clustering
statistics on all scales. The distinct spatial distribution of the galaxies
in Population 2 which were assigned fibers can be properly accounted for by
their angular completeness $N_2'/N_2$ in each individual tile sector. We
explain the theoretical basis for our method and extensively test it on
realistic mock catalogs. We demonstrate that both the projected and the 3D
redshift-space 2PCFs can all be very well recovered, assuming that the
recovered Population 2 galaxies are representative. With that assumption, the
correction method is accurate and unbiased and limited only by sample
variance.

In the real observation, the density variations between overlap and
non-overlap regions and the non-random selection of the Population 2 galaxies
in higher colliding groups are the main potential systematic effects. While
the density effect can be largely ignored as we demonstrate in the mocks, the
influence of higher colliding groups should be carefully taken into account
in order to provide a reasonable estimate of the true 2PCFs. We proposed a
simple fix to the missing close pairs caused by such colliding groups. The
remaining systematic effect on small scale 2PCFs from such a method is
expected to be lower than $\sim5\%$. The exact values of the systematic
errors are hard to obtain, given that we are limited by the sample variance
on small scales. By shifting the plate placement of the tiling geometry, we
find that the deviation from the true 2PCF based on our correction method is
likely to be below $3\%$. To be conservative, we quote $5\%$ as an upper
limit of the systematic errors for our method.

We also contrast our method to the commonly used nearest-neighbor and angular
corrections. While these approximations work well for specific statistics on
some scales, they are generally not accurate enough to give an unbiased
estimate over all scales. For the nearest-neighbor correction, assigning the
collided galaxy the redshift of its neighbor mainly influences the
line-of-sight separation $\pi$, while $r_p$ is only changed because of the
non-plane-parallel effect. For projected statistics like $w_p(r_p)$, on
scales larger than the fiber-collision scale, the pair counts are dominated
by non-collided galaxies, and the nearest neighbor correction works extremely
well. On small scales, however, the pair counts are dominated by collided
galaxies. The nearest neighbor redshift assignment leads to an overestimate
of the number of pairs within the projected separation $\pi_{\rm max}$ and
thus an overestimate of $w_p(r_p)$, which causes the nearest neighbor
correction to fail. For $\xi(s)$, the nearest-neighbor correction fails below
and above the fiber-collision scale, since having the correct line-of-sight
separation is more crucial,  and the method does worse on all scales smaller
than $15\mpchi$. The nearest-neighbor correction, however, is still a good
estimate on very large scales.

The angular correction method provides a better estimate on small scales than
the nearest-neighbor one, but it is still only an approximation. By
definition, the angular 2PCF $w(\theta)$ is obtained via projecting over the
whole line-of-sight depth of the survey, while the projected 2PCF $w_p(r_p)$
is obtained from projection within line-of-sight separation of $\pi_{\rm
max}$. Therefore, $w(\theta)$ and $w_p(r_p)$ are not expected to be exactly
the same. On small scales, the effect of the difference in projection depth
is negligible and the angular correction method works reasonably well, while
on large scales it is no longer the case. In addition, for a survey with
large line-of-sight depth, the mapping from $\theta$ onto $r_p$ is not unique
because of the non-plane-parallel effect, further complicating the
correspondence between $w(\theta)$ and $w_p(r_p)$ and making the angular
correction method less accurate. We find that our correction is generally
superior, better theoretically motivated and more broadly applicable,
especially in its power to recover the 3D correlation functions. Since the
angular correction relies on the measured redshifts of the galaxies in the
catalog, the systematic effects of the density variation and higher colliding
group also affect the accuracy of the angular correction on small scales.

Our correction method can be directly applied to flux-limited survey samples.
When dealing with a subsample of galaxies in certain redshift, luminosity or
color bins, e.g., volume-limited samples, the determination of $D_1$ and
$D_2$ populations is not as obvious, since we do not have redshifts for the
missing galaxies. But as the two populations are only separated based on
their angular distribution, which is independent of redshift, luminosity or
color, the determination of $D_1$ and $D_2$ populations should be made in the
parent full sample, where all the galaxies satisfy the same selection
criteria. The $N_2'/N_2$ fractions in each sector are then also determined in
the parent sample. The total number of $N_2$ galaxies in each subsample is
calculated by summing up all the expected $D_2$ galaxies in each sector,
$N_{2,{\rm sec}}=N_{2',{\rm sec}}N_2/N_2'$. Based on a test of subsamples of
galaxies in different color bins in CMASS DR9, we find that the color
distribution of $D_2'$ galaxies determined in this way is the same as that of
the $D_2$ galaxies, supporting our assertion that the division into the $D_1$
and $D_2$ populations is independent of the physical properties of the
galaxies.

In developing the new method, we have considered many other alternative
corrections, such as using photometric redshifts, imposing fiber-collisions
on the random catalog, applying different weights on the galaxies. We have
also considered different variants in our definition of Populations 1 and 2.
However, none of the many variants we have tried succeeded in robustly
recovering the underlying clustering. Our new method, rooted in solid
theoretical ground, proves to work successfully.

For the early CMASS-like mock catalogs used in this work ($\sim 50,000$
galaxies in a volume of $\sim 0.16 h^{-3}{\rm Gpc}^3$) with imposed tiling,
our correction method reaches a statistical accuracy of $\sim 6\%$ at
$r_p\sim0.1\mpchi$ and $\sim 2.5\%$ at $r_p\sim30\mpchi$ for $w_p(r_p)$, and
systematic errors of $\lesssim 5\%$ on small scales. The statistical errors
are essentially caused by the sample variance, and thus will be reduced for
larger volumes, scaling down by roughly the square root of the volume. We
have verified that this scaling law holds when using subsamples of smaller
volume, and the fluctuation around unity of the ratio of 2PCFs averaged over
the 40 mocks in Figure~\ref{fig:wptile} also shows that it scales down
accordingly. The current SDSS-III BOSS DR9 sample already covers an area of
about $3500\deg^2$, roughly $6$ times larger than our mocks, so our
statistical correction error will be $2.4$ times smaller when applied to this
sample. The final survey will cover about $10000\deg^2$, and the correction
error in $w_p$ from sample variance is expected to be $\sim 1.5\%$ at
$r_p\sim0.1\mpchi$ and $0.6\%$ at $r_p\sim30\mpchi$. The residual systematics
may also scale down somewhat with the decreased sample variance, and we plan
to develop a better treatment of them in future work.

The method thus enables more accurate measurements of galaxy clustering
statistics on small and intermediate scales. In particular, it will enable us
to reliably extend the measurements to smaller scales than obtained before
and to recover the full 3D redshift-space correlation functions on small
scales. This will allow a better probe of the distribution of galaxies within
halos. HOD modeling of these new measurements will provide new constraints on
the spatial and velocity distributions of galaxies inside halos.  Measurement
and modeling of redshift-space distortions in both the linear and non-linear
regimes will also improve constraints on cosmological parameters. These
applications will be explored in future work. We note, however, that as our
method is associated with increased sample variance, due to the limited sky
coverage of $D_2'$ galaxies, it might not be ideal for very large scale
clustering measurements, such as those performed for measuring the baryon
acoustic oscillation signature \citep{BAO}. In such cases, other methods such
as the nearest neighbor correction can be considered.

The new method will allow for reliable galaxy clustering measurements in
current and future surveys. While we were motivated by upcoming measurements
in the SDSS-III BOSS survey, and tested the method on corresponding mock
catalogs, the method can be applied to any fiber-fed large surveys such as
the SDSS-I and II, 2dFGRS, and planned BigBOSS \citep{BigBOSS}. We have
focused here on two-point auto-correlation functions, but our methodology is
more broadly applicable to other related statistics. It can be easily
extended to cross-correlation functions (see the Appendix), which would not
suffer from the systematic effect of higher-order colliding groups if the two
samples of targets come from different surveys. The correction method can be
further generalized to high-order statistics, e.g., the three-point
correlation function (\citealt{Jing04,Kayo04,McBride11}; H.\ Guo et al. 2012,
in prep.), with the effect of collision groups treated more accurately. The
method can certainly be helpful in accurately measuring statistics in Fourier
space as well, such as the power spectrum and bispectrum of galaxies.

\acknowledgments

We thank the anonymous referee for insightful suggestions that significantly
improved the paper. We thank Cameron McBride for providing the LasDamas mocks
and Chris Mihos for providing access to the High Performance Research
Computing Cluster at Case Western Reserve University. We also thank Will
Percival, Ariel Sanchez, Ashley Ross, Martin White, Andreas Berlind, Michael
Blanton, Daniel Eisenstein, Shirley Ho, Yipeng Jing, Cheng Li, Nikhil
Padmanabhan, John Parejko, Jeremy Tinker, David Wake and David Weinberg for
stimulating discussions and the BOSS galaxy-clustering working group as a
whole for support and encouragement. We thank the LasDamas project for making
their mock catalogs available. This work is supported by NSF grant
AST-0907947.

\appendix

Our new method for fiber-collision correction is also valuable for the
measurement of cross-correlation functions. For cross-correlation functions,
the key equation (\ref{eqn:xid'}) is revised accordingly. When
cross-correlating two samples $\it{a}$ and $\it{b}$, both are divided into
$\it{Populations\ 1}$ and $\it{2}$. The decomposition equation becomes
%%%%%%%%%%%%%%%%%%%%%%%%%%%%%%%%%%%%%%%%%%%%%%%%%%
\begin{eqnarray}
N_aN_b\ \xi&=&N_{a_1}N_{b_1}\xi_{a_1b_1}+N_{a_1}N_{b_2}\xi_{a_1b_2}+
N_{a_2}N_{b_1}\xi_{a_2b_1}+N_{a_2}N_{b_2}\xi_{a_2b_2}\\
&\approx&N_{a_1}N_{b_1}\xi_{a_1b_1}+N_{a_1}N_{b_2}\xi_{a_1b'_2}+
N_{a_2}N_{b_1}\xi_{a'_2b_1}+N_{a_2}N_{b_2}\xi_{a'_2b'_2},\label{eqn:cross}
\end{eqnarray}
%%%%%%%%%%%%%%%%%%%%%%%%%%%%%%%%%%%%%%%%%%%%%%%%%%
where $N_{a_1}(N_{b_1})$ and $N_{a_2}(N_{b_2})$ are the numbers of galaxies
in Populations $1$ and $2$ in sample $a(b)$, and $a'_2$ and $b'_2$ denote the
subsets of $a_2$ and $b_2$ for which redshifts were obtained, analogous to
$D_2'$ and $D_2$.

\end{document}